\documentclass[sigconf]{acmart}

\usepackage{booktabs} 
\usepackage{float}
\usepackage{subcaption}

\acmDOI{}

\acmISBN{}

\acmConference[]{} 
\acmYear{}
\copyrightyear{}

\acmArticle{4}
\acmPrice{15.00}

\begin{document}
\title{Improving S\&P stock prediction with time series stock similarity}

\author{Lior Sidi}
\orcid{1234-5678-9012}
\affiliation{%
  \institution{Department of Software and Information System Engineering\\ Ben-Gurion University of the Negev}
  \city{Israel}
}
\email{liorsidisc@post.bgu.ac.il}

\renewcommand{\shortauthors}{L. Sidi}

\begin{abstract}
Stock market prediction with forecasting algorithms is a popular topic these days where most of the forecasting algorithms train only on data collected on a particular stock. In this paper, we enriched the stock data with related stocks just as a professional trader would have done to improve the stock prediction models. We tested five different similarities functions and found co-integration similarity to have the best improvement on the prediction model. We evaluate the models on seven S\&P stocks from various industries over five years period. The prediction model we trained on similar stocks had significantly better results with 0.55 mean accuracy, and 19.782 profit compare to the state of the art model with an accuracy of 0.52  and profit of 6.6.


\end{abstract}

%
%



\keywords{ACM proceedings, Time series similarity, Stock predictions}

\maketitle

\section{Introduction} \label{introduction}
Prediction of stock price or any financial equity is well-investigated subject by many researchers \cite{Finnie2010FinancialSurvey}, traders, and hedge funds. In an entire algotrading framework, the stock prediction component collects information from different sources, such as market trading and news. The components' goal is to feed the strategy component with feed on the following price values. The strategy component is responsible for digesting the information regarding the current trader position, risk parameters, losses, and the price prediction to set actions for buying or selling particular finance equity.  

Similarity analysis on time series data in the finance domain is used widely to cluster different equities into domains for manual exploration \cite{Tian2016TheClustering} \cite{Aghabozorgi2015Time-seriesReview} but also to identify correlated stocks for trading strategy \cite{Aghabozorgi2014StockMethod} and for stock recommendation \cite{Nair2017ClusteringStudy}. 

In this research, we investigate if a prediction model improves by adding similar stock during training and in prediction. Our two main research questions are "is the enhancement of similar stock improve a model performance" and "which similarity configuration improves the model the most?" 

From the extensive literature overview we applied, We believe that we are the first to evaluate how similar stocks enchantments can improve the prediction of a stock price. We prepared a back-testing framework to train and optimize prediction models with different time series processing, segmentation, and modeling. In the Methods chapter, we describe the back-testing pipeline and explain the different similarity functions we apply for finding the most similar stocks. In the Experiment setup chapter, we explain how we optimize and evaluate the back-testing process on S\&P stocks. Finally, in the experiment results chapter, we compare between models that trained on similar stocks with models that trained only on the target stock or random stocks. The results show a clear advantage of models trained on similar stocks with a short horizon period (the next day) with a mean profit of 19.87 and a mean accuracy of 0.55.

\section{Related work}\label{related}

In this research, we aim to improve stock prediction with a hybrid approach that combines stock similarity and classification. This section starts with an overview of the stock data representation and evaluation. We review different appliances of stock similarities and clustering techniques. 

\subsection{Stock time series overview}
\citeauthor{Cavalcante2016ComputationalDirections} describes a two parts framework for financial trading forecasting; the first part deals with conventional forecasting aspects such as data preparation, algorithm choosing, model training, and accuracy evaluation. The second part is responsible for financial forecasting aspects such as trading strategy and then profit evaluation.

\subsubsection{Stock time series data\newline}
A standard financial data usually consist of aggregated data of the stock price for a certain period. The aggregations are usually high price, low price, opening price, closing price, trade volume, trading amount. Many papers also extract known technical indicators to identify trends and momentum in the stock price \cite{Vanstone2009AnNetworks}. Table \ref{table:indicators} cover the most important technical indicators. 

A critical aspect in the data preparation is the prediction period, a short horizon period such as one day, one week, and one month is more suitable for financial prediction with technical indicators \cite{Evans2013UtilizingSpeculation}.

\begin{table*}
  \caption{Technical indicators for stock time series}
  \label{table:indicators}
  \begin{tabular}{ll}
    \toprule
    Technical indicator & description \\
    \midrule
    relative strength index (RSI) & holds the magnitude of recent \\
                                  & gains and losses over a specified\\
                                  & time period \\
       rate of change (ROC)           & estimating the speed of change \\
                                    & in a price \\
    moving average convergence  & accumulating the relationship\\
    / divergence (MACD)         & between two moving averages of prices \\
    Sharpe ratio                 & calculating the risk of a certain \\
                                & period by subtracting the profits \\
                                & with the standard deviation\\
    
    \bottomrule
  \end{tabular}
\end{table*}

\subsubsection{Segmentation\newline}
Time series representation and segmentation are a major part of many stock time-series tasks. The goal is to reduce the dimensionality and complexity of the data and enable identification of technical patterns, clustering, or prediction \cite{Cavalcante2016ComputationalDirections}. 

\subsubsection{Prediction\newline}
Forecasting proven to be well-suited for financial data modeling, sophisticated machine learning (ML) models such as artificial neural networks, SVM and genetic programming showed state of the art results in the field.\cite{Hu2015ApplicationReview} \cite{Evans2013UtilizingSpeculation} \cite{Vanstone2009AnNetworks}.

\citeauthor{Finnie2010FinancialSurvey} surveyed different techniques for time series forecast on financial data; they differentiate between machine learning technique, forecasting period, and the input variables. They summarized with the noticed that the Artificial Neural Networks (ANNs) is a dominant machine learning technique. Nevertheless, \citeauthor{Gerlein2016EvaluatingApproach} demonstrates that also simple ML models such as decision tree, logistic regression, nearest neighbor, and Naive Bayes showed good results as well. Therefore using simple forecast algorithms can be a good benchmark to evaluate different representation methods and enrichments.

\subsubsection{Evaluation\newline}
Evaluating a stock time series predictor involves two types of metrics, the first is a conventional evaluation of the predictor with the accuracy measures such as mean absolute error, mean absolute percentage error, and root mean square error. The second is money evaluation, which evaluates the profit for a certain trading strategy \cite{Cavalcante2016ComputationalDirections}.

One of the most popular strategies for evaluation is a Buy \& Hold strategy \cite{Finnie2010FinancialSurvey}. The strategy simply buys a stock that predicted to go up and sell it otherwise. Still, any strategy shall apply a risk control mechanism such as stop loss\cite{Chande1997BeyondAnalysis}.

\subsection{Similarity and clustering on stock time series}
Clustering in time series stock data serves many goals such as portfolios balancing, patterns discovery, risk reduction, finding similar companies, prediction, and recommendation. 
Applying a clustering model requires three key components: clustering algorithm, similarity definition, and evaluation method \cite{WarrenLiao2005ClusteringSurvey}.

\citeauthor{Keogh2002OnBenchmarks} address two types of clustering time series, the first is whole time-series clustering to cluster set of individual time series by their similarity, and the second is subsequence clustering to extract subsequences per time series such as sliding window.
\citeauthor{Aghabozorgi2015Time-seriesReview} also adds time point clustering to cluster the points values in the time series. 

The similarity and the clustering are highly affected by the segmentation and representation method applied, \citeauthor{Keogh2002OnBenchmarks} boldly claim that the clustering of time-series subsequences with time window is meaningless unless only the significant motifs are considered.

In the rest of this section, we will explore the recent appliances of clustering on financial time series data per type of similarity distances.

\subsubsection{Numeric distance\newline}
\citeauthor{Aghabozorgi2014StockMethod} used different clustering methods to categorize companies based on their stock data similarity; they used basic Euclidean distance to find similarity in time points in stock data. Because Euclidean distance is not capable of identifying trends shapeliness they used Dynamic Time Warping (DTW) distance to find similarity in the stock data shape, DTW first introduced in the 1960s and still show similar results to more advance methods \cite{Ding2008QueryingMeasures}. In general, DTW deals with unequal length and solves the local shift problem in the time series to find similar shapes between time series in different time phase axis. 

\citeauthor{Wang2012SimilarityTree} applied DTW on foreign exchange (FX) market and use minimal spanning tree (MST) and hierarchical tree (HT) to cluster different currencies together. They strongly claimed that the usage of the Pearson correlation coefficient (PCC) is not suitable for FX time series data because it is not robust to outliers and must have homogeneous, synchronous, and equal length samples.

\citeauthor{Jeon2017PatternData} searched for similar patterns in historical stock data with DTW and stepwise regression feature selection to improve predictions. The selected data set is used to train an artificial neural network (ANN). They evaluate their predictor with root mean square error (RMSE) and new evaluation that represent the target value with SAX and apply Jaro-Winkler similarity.

\citeauthor{Caiado2007AEvidence} used generalize autoregressive conditional heteroskedasticity (GARCH) models to estimate the distance between stock time series volatilities. They used hierarchical clustering and multidimensional scaling technique to differentiate geographical stock markets. GARCH model assumes that the conditional variance is dependent on a past linear volatility model; the GARCH model is lean with parameters and provides a good representation of volatility for a variety of processes. Their distance formulation considers the time series GARCH measurement combined with the sum of the series covariance-vector-estimation.

\subsubsection{Symbolic distance\newline}
\citeauthor{Soon2007AnData} compared the numeric and symbolic representation for stock data similarity. For numeric representation, they use the original data with Euclidean distance, and for symbolic representation, they used UP, DOWN, and SAME symbols with a number of matching symbols as distance.
They found that opening, closing, highest, and lowest prices of the stock are able to produce consistent results in similarity and demonstrate that under the representation and distances described above, the numeric distance was more consistent then symbolic distance.

\citeauthor{Aghabozorgi2014StockMethod} used Symbolic ApproXimation Aggregation (SAX) representation for dimensionality reduction, SAX method discrete stock continues representation with symbols per static data segment; they used a k-Modes algorithm that suites categorical data. For the SAX distance measurement, they develop APXDIST instead of MINDIST distance for symbolic distance \cite{Liu2009ResearchData} because MINDIST considers the neighbor symbols as zero, APXDIST distance also considering the global minimum and maximum symbols in the sequence.

\citeauthor{BrancoPatternSAX} combined SAX and Shape Description Alphabet (SDA) representation with a genetic algorithm to generate buy and sell signals. SDA representation calculates the amplitude difference between two adjacent points and represents it as a symbol. SAX is not capable of identifying the difference between segments that have the same average value, whereas the SDA identifies trends and relation between adjacent points. For SAX, they use MINDIST, and for SDA, they use simple numeric subtraction between the relative representations.  

\citeauthor{Tamura2016TimeEvaluation} conducted time series classification based on SAX representation with Moving average convergence divergence (MACD) Histogram and applied one nearest neighbor (1NN) with extended Levenshtein distance that suites strings with SAX representation. MACD captures the velocity and the acceleration of the time series and is used widely in the financial domain, MACD calculates the difference between two exponential moving averages (EMA) with different window size. \citeauthor{Tamura2016TimeEvaluation} used SAX to represent the original values and MACD values, and then they combined the values in an alternates order. 

\section{Methods}\label{methods}
We developed a workflow that prepares the stocks data and manage the back-testing process, the workflow's generic implementation allows the evaluation of different methods configurations. 

In this chapter we will describe the workflow's stages and methods we apply in this research, the workflow code is written in Python and is available on GitHub: \url{https://github.com/liorsidi/StockSimilarity}. The workflow pipeline has four stages: Prepare fold data, process data, similar stocks enhancement, and stock prediction. 

\begin{figure}[h]
\centering\includegraphics[width=1\linewidth]{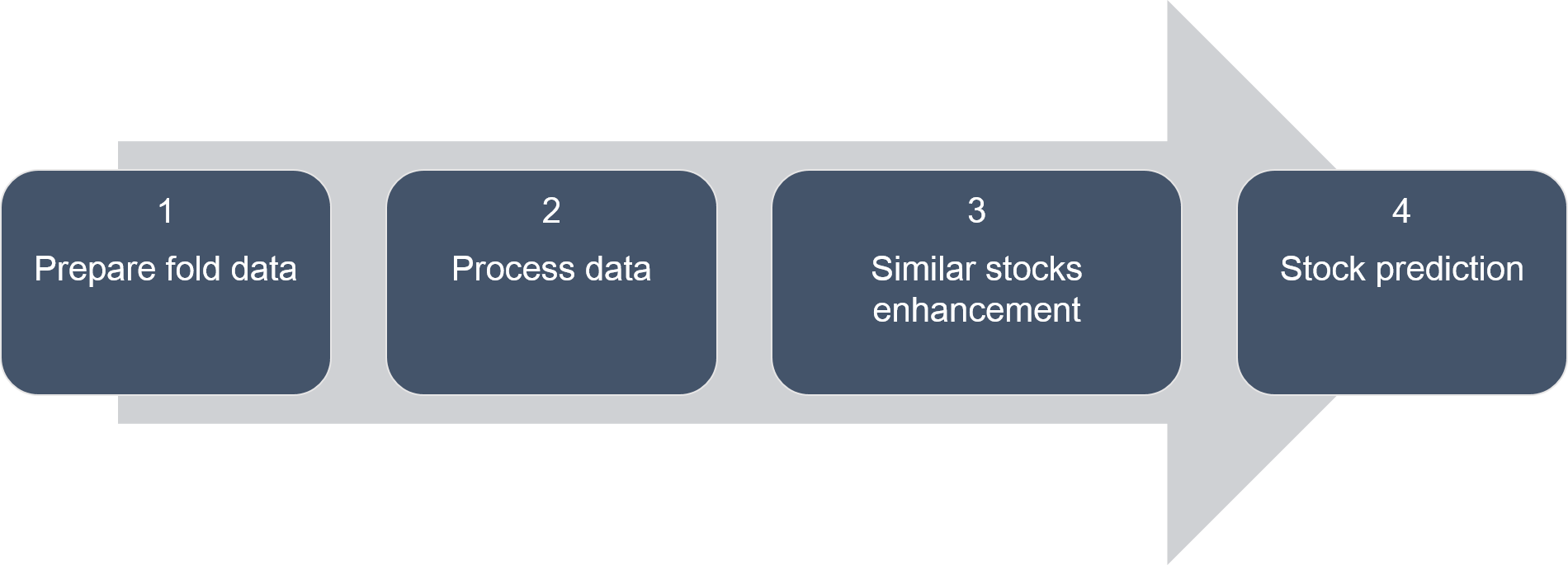}
\caption{Workflow pipeline stages}
\label{fig:pipeline}
\end{figure}

\begin{figure*}[h]
\centering\includegraphics[width=0.8\linewidth]{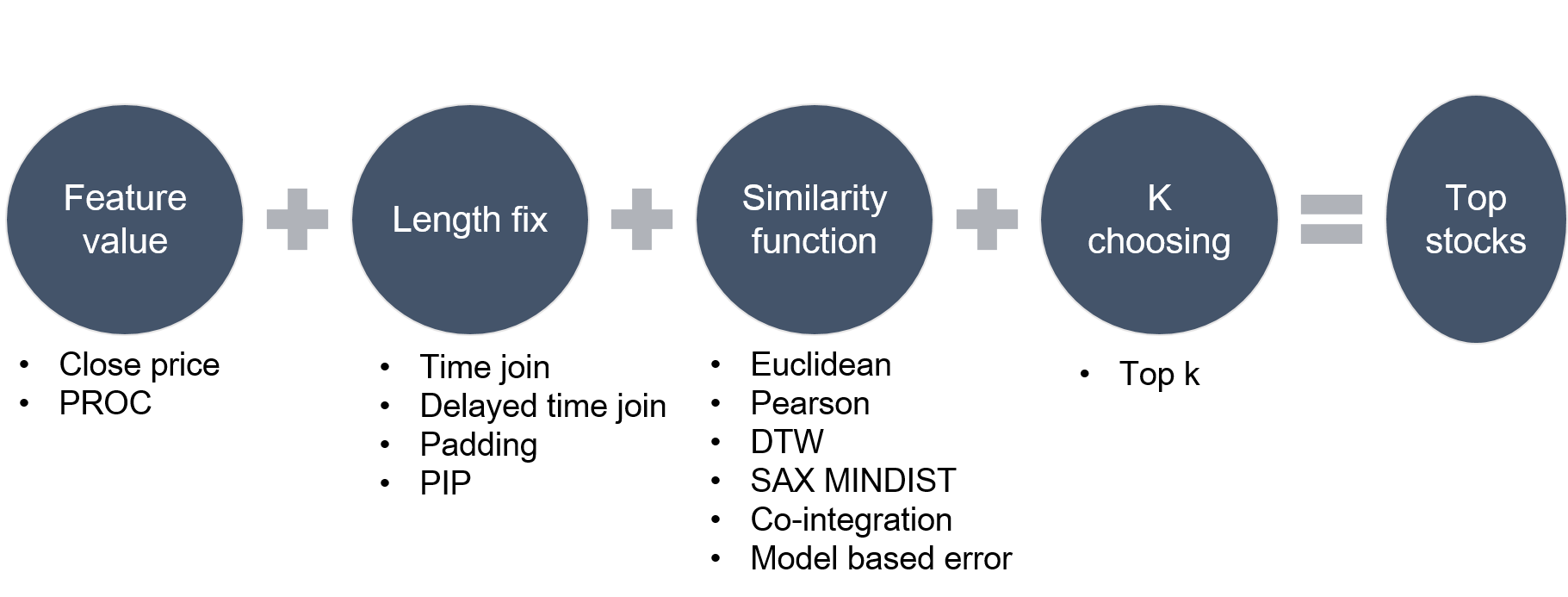}
\caption{The different configurations for stock similarity}
\label{fig:topstock}
\end{figure*}

\subsection{Folds preparation}
In this stage, the process split the data to n folds; the split separates the data to n equal width folds. Each fold contains a train data, a fold before the split, and a test data, a fold after the split. 

in order to keep the model relevant to the test data, We select only half of the test data near a split point, 

\subsection{Data processing}
The processing stage is responsible for manipulating, modeling, and feature extraction of the stock data. The actions performed by this component are normalization (standardization), financial feature extraction (MACD, RSI, Price rate of change, volume, open-close difference, and trading volume), data segmentation (SAX or PCA) and data modeling (time points or time windows). 

The component is also responsible for training the normalization and segmentation processes on each train fold data and apply it to the relevant test fold. Furthermore, this stage differentiates between univariate modeling with only one value and multivariate modeling with all the financial features. We apply time window modeling only for the univariate to reduce the size of dimensionality and eliminate overfitting. We also did not apply normalization on some financial features because their original raw values already normalized between ranges.

\subsubsection{Standard normalization}
The standardization phase normalizes the financial data. The normalization is possible in the stock prediction scenario because of the values of random behavior. The min-max normalization is not advised because the stock values are not limited to a certain price range, and the prices can change dramatically. Therefore, we normalize the stock prices with elementary standardization function: 
\[ z = 
    \dfrac{x - mean(X)}{std(X)}
\]

\subsubsection{Data modeling}
Time series data modeling can take place in various ways; each modeling represents the prediction record differently and exposes different information regarding the instance and its context.

\begin{itemize}
  \item \textbf{Timepoint} - Addresses each time point separately without data of previous time points. Timepoint representation allows adding complexity like feature extraction or combining similar stocks values at the same timepoint. 
  
Nevertheless, timepoint approach has several limitations, the first is a significant reduction of data when joining similar stocks on the same time point due to missing data, and the second limitation is the lack of contextual information from recent points. Therefore, adding financial features that capture previous information such as exponential windows might be more beneficial. 
  
  \item \textbf{Time Window} - Represents each instance as a window of adjacent time points, this modeling allows the model to find a relation between adjacent points. Still, this modeling enlarges the instance and can harm the ability to add more features and stocks. 
  
In our implementation, we extract windows for each stock separately and do not join their values as applied in the time point modeling. On the one hand, the similar stocks data is not used in prediction, only in training. But on the other hand, the training data increases dramatically with similar stocks and improves the models. \end{itemize}
 
\subsubsection{Data segmentation}
\begin{itemize}
 \item \textbf{Principals component analysis (PCA)} - creates a smaller representation of the dataset while maintaining its variance using eigenvector decomposition on the data covariance. 
 
The PCA produces principal components (PCs), which are a linear combination of different features that acts as a new attribute. PCA is a common tool for data exploration and allows good reasoning of the data variance.
 
In our experiment, we set the PCA to produce 3 PCs from the entire data set to act as new attributes.

 \item \textbf{Symbolic ApproXimation Aggregation (SAX)} - A dimensionality reduction technique, allows distance measures to be defined on the symbolic approach that lower bounds euclidean distance \cite{Lin2003}. 
 
SAX involves performing two stages on the data: first, it transforms the original time-series into the appropriate piecewise aggregate approximation (PAA) representation. 
 
The PAA representation divides the series to parts (according to the given output length) and calculates each interval's mean value. Later it converts the PAA data into a string after normalization according to the given alphabet size. In our experiment, we use SAX representation while keeping the same word size as the origin.
\end{itemize}

\subsection{Similar stocks enhancement}
The stock enhancement component applies the different functions for measuring the stock similarity. In figure ~\ref{fig:topstock}, we describe the necessary configurations, the value for calculating the similarity, the fix the length function, the similarity functions, and lastly, the k top stocks are chosen. 

As explained in the previous data modeling section, The combination of similar stocks depends on the data modeling approach.

\subsubsection{length fixing}
Each instance (stock or equity) may miss different time points due to system error, vacation days, or stock splitting.

When computing the distances between time-series, it's important to correlate them to have the same length size, in order to do this, we implemented and examined the following fixing methods:
\begin{itemize}
  \item \textbf{Time join} - a basic correlation between stocks, if one stock is missing a time point while the other is not the time point is eliminated (equivalent to inner join in SQL) this fixing is the most popular but may reduce the data substantially.
  \item \textbf{Delayed time join} - stock values are pushed t times points backward (delay), this correlation meant to identify if one stock indicated future behavior of the other one. 
  \item \textbf{Padding} - basic padding fixing technique that adds a duplicate value at the beginning of the shorter series.
  \item \textbf{Perceptually important points (PIP)} - select the most important points in a series with the following steps: the first and the last points are set as PIP's. Then, the third PIP will be the point with the maximum distance to the first two PIP's. The fourth PIP will be the point with the maximum distance between two adjacent PIPS, the algorithm finish when achieving a predefined number of points. 
  
In our experiment, We use PIP on each stock to find important time points (10 percent of original length), and then we combined both PIPs and correlated the stock time points.
\end{itemize}

\subsubsection{Similarity functions}

\begin{figure}[h]
\centering\includegraphics[width=1\linewidth]{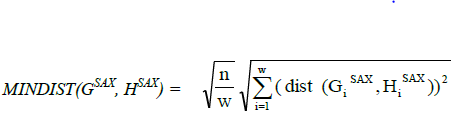}
\caption{MINDIST Equation - n is the number of window points, w the number of segments, Gi the value number i of series G transformed by the SAX method, and the Hi the value number i of series H transformed by the SAX method.
}
\label{fig:MINDIST}
\end{figure}

\begin{itemize}
  \item \textbf{Euclidean distance} - A common indicator that measures the dissimilarity between time series comparing the observations at the exact same time. The Euclidean distance is a square root of the sum of the squared differences of each pair of corresponding points. 
  
The main limitation of this measure is its inability to identify shifting and trends in the data.
  
  \item \textbf{Pearson correlation coefficient } - A known measure of the linear correlation between two vectors, the coefficient is calculated by dividing the two series covariance with theirs standard deviation product, the correlation value range is between -1 and 1 for negative and positive correlation.  
  
Pearson has two major limitations regarding stock price correlation. The first is that it assumes stationary behavior, and the second is that it cannot deal with non-linear behavior between series. 
  
  \item \textbf{Dynamic Time Warping (DTW)} - A template matching algorithm in pattern recognition, DTW, which can align sequences that vary in time or speed. 
  
DTW is an old technique but still very relevant in financial similarity. In our experiment, we used Python's implementation of DTW, based on Euclidean distance. 
\item \textbf{MINDIST} - A distance computation of SAX representation, the MINDIST formula defined by \cite{Lin2003} and explained in figure ~\ref{fig:MINDIST}. The main limitation of MINDIST in stock price series, as mentioned by \citeauthor{Liu2009ResearchData}, is that it does not address adjacent change in values. To fix this, we tested MINDIST (and all other similarity functions as well) on the price rate of change (PROC) to identify a high increment of adjacent change.
\item \textbf{Co-integration} - A statistical feature between multiple non-stationary time series, co-integration checks if there is a parameter that it's multiplication with one of the time series resolve with a constant spread between the non-stationary series.  

Stock prices are not necessarily stationary because their mean and standard deviation may change over time. Co-integration is used widely to compare similarity between stocks and may state that there is some relation between them \cite{Alexander2003EquityEfficiency}. 

For testing series co-integration, we use the co-integration Python implementation "stattools" library that test for co-integration behavior with Engle-Granger two-step co-integration test. We used the test P-value as a similarity measurement between the two series, a low p-value of the test means that the series are co-integrates.
\end{itemize}

\subsection{Stock prediction}
The prediction component trains a regressor or a classifier model with the relevant algorithm. The model goal is to predict if the stock value will increase or decrease to a certain horizon. 

\begin{figure}[h]
\centering\includegraphics[width=0.8\linewidth]{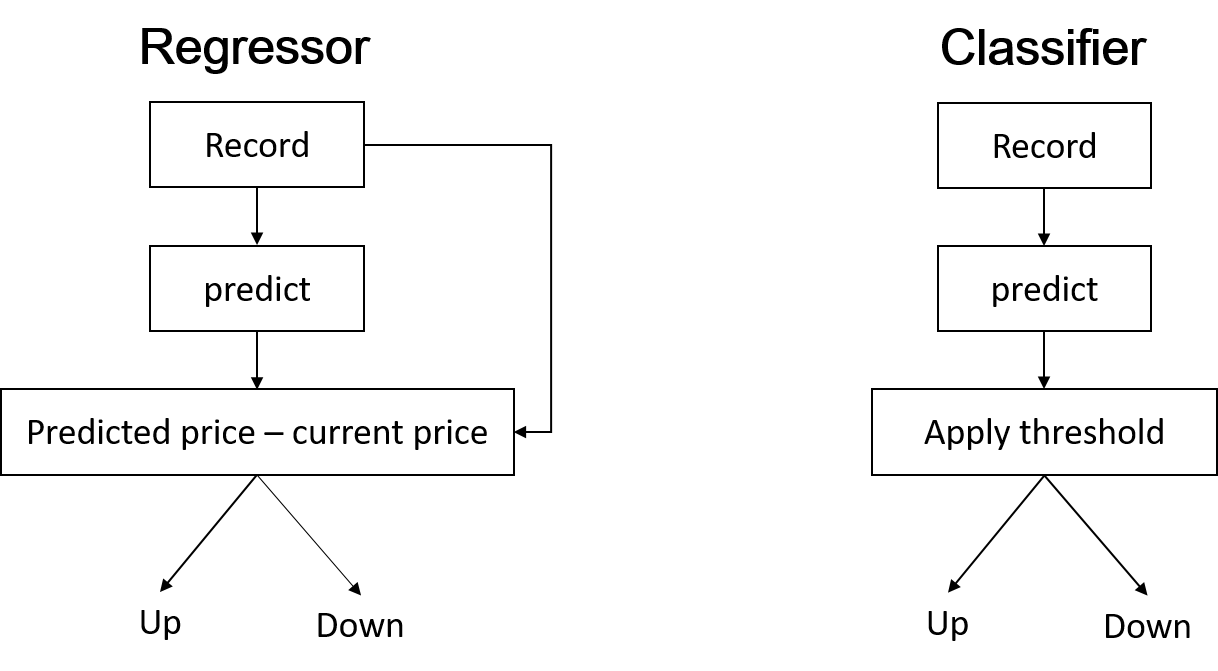}
\caption{The training and appliance pipeline for classifiers and regressors models}
\label{fig:clasvsreg}
\end{figure}

Figure ~\ref{fig:clasvsreg} explains how the system train and apply the classifier or the regressor model. For the classifiers, the system train on the two classes in the traditional way. The regressors are trained to predict the next price value, but in prediction, the predicted value is a binary value for an increase or decrease in the price value.

In our experiment, we choose to evaluate two ensemble algorithms that showed good results in the stock prediction domain: Random Forest and Gradient Boosting Tree. Both models have a classification and regression implementation in Scikit-learn (Python library).

\begin{itemize}
\item \textbf{Random Forest} - Train t decision trees on different features and, in prediction, apply a majority voting on the results from all trees.
\item \textbf{gradient boosting trees} - Train a chain of decision trees where each tree tries to predict the error of the previous decision tree, the model has a learning rate for summing the values from the tree chains. 
\end{itemize}

\section{Experiment setup}\label{setup}
\begin{figure*}[h]
\centering\includegraphics[width=0.8\linewidth]{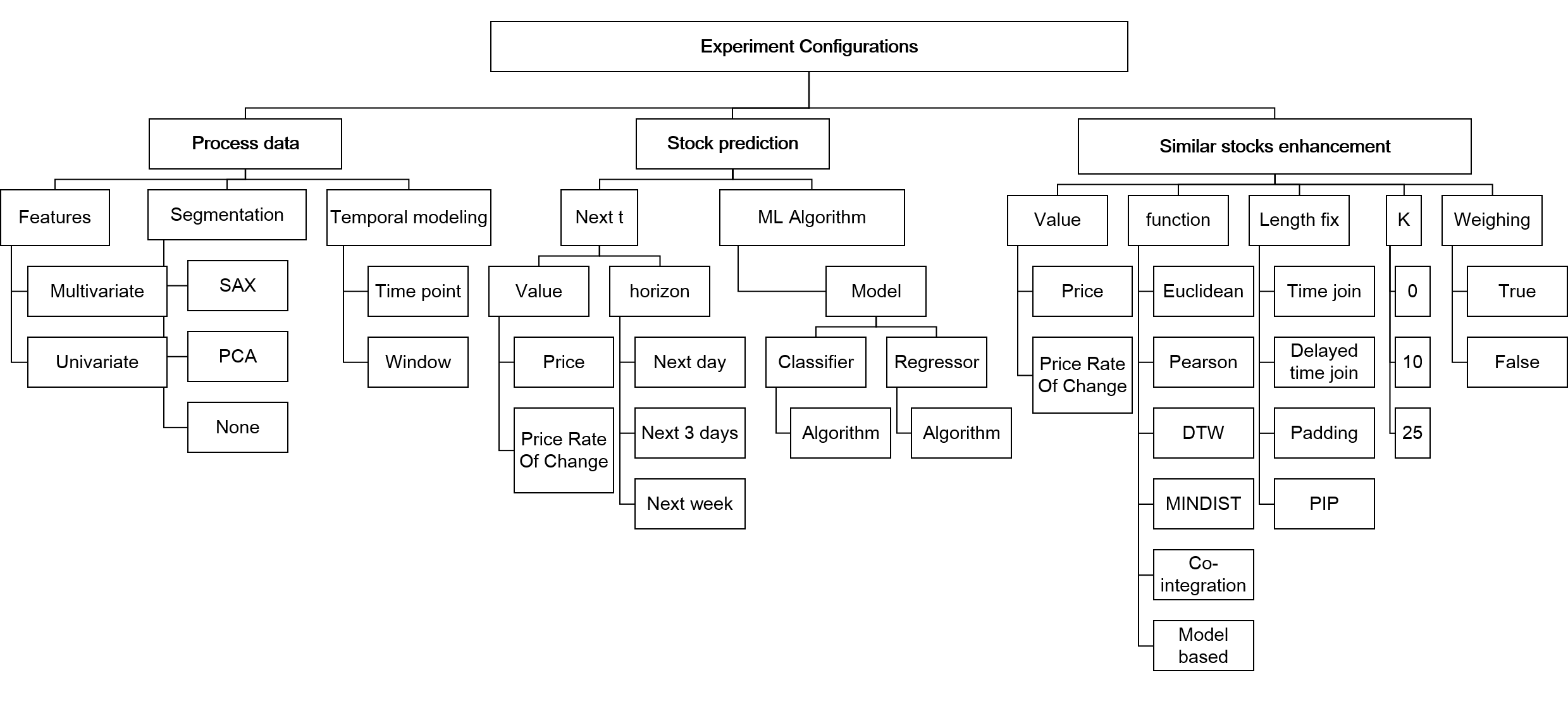}
\caption{A configuration tree of all the setup to be optimize and evaluate in workflow pipeline}
\label{fig:config}
\end{figure*}

In order to evaluate if stocks similarity improves a baseline model, we conduct two-step experiments (back-testing) to evaluate different types of configurations. The first experiment goal is to come up with a processing pipeline and a baseline model. The second experiment is to evaluate how different stock similarity functions influence the baseline model. 
In figure ~\ref{fig:config}, we mapped the different configuration parameters the back-testing process will evaluate. 

Our dataset contains daily historical data for all the S\&P (Standard \& Poor) 500 stock market index companies from 2012 to 2017. The features given are date, open price, closing price, highest price, lowest price, volume, and the short name of the stock. The S\&P is an American stock index of the largest companies listed in NYSE or NASDAQ, maintained by S\&P Dow Jones Indices. It covers about 80 percent of the American equity market by capitalization.

We apply the evaluation process on stocks from different industries: Consumer (Disney - DIS, Coca Cola KO), Health (Johnson and Johnson - JNJ), Industrial (General electric - GE , 3M - MMM), Information technology (Google - GOOGL) and Financial (JP Morgan - JPM). The validation folds are set to five and prepared for each stock separately.

\subsection{Experiment 1 - processing model evaluation}
The experiment's goal is to evaluate the basic processing and prediction model parameters to set a baseline model and processing configurations. The baseline settings are set in the next experiments to evaluate the similarity enhancement, rather them model and processing tuning. 

For processing, the experiment evaluates features (univariate or multivariate), segmentation methods (SAX, PCA, or raw values), temporal modeling (time points, or windows size 5 or 10). For stock prediction, the experiment evaluates the following configurations: prediction value (close price or price rate of change), horizon (next day, next three days or next week), and weighing instances per stock (applied only for the Euclidean similarity models).

To identify if stock similarity enhancements improve a baseline model we do not need to focus on improving the models with endless parameters tuning, but set recommended parameters to reduce the complexity of the experiments, the models recommended configuration: Random forest with 100 trees and Gradient boosting with 0.02 learning rate.
The first experiment also applies two basic stock similarity configurations: Euclidean similarity function on the price value with ten similar stock compared to no enrichment of similar stocks. 

\subsection{Experiment 2 - Enhancement similarities evaluation}
The second experiment evaluates the improvement each similarity function parameters have on the baseline models defined in the first experiment. 

For similarity parameters, the experiment evaluates similarity functions (co-integration, DTW, Euclidean, Pearson, and SAX), length fixing functions, size of k similar stocks (10, 25, 50), similar value to compare (Close price or price rate of change).

The experiment will compare the best-enhanced model with the best non-enhanced model from the first experiment and with a model that randomly choose stocks for enhancement. The second comparison goal is to evaluate if the model's improvement is due to similarity enhancement and not due to general stock enrichment.

\section{Results}\label{results}
The evaluation metrics are accuracy score and F1 score; we calculate each metric per class (increase/decrease) and average it to one score. To evaluate the model profit, we implement a simple Buy \& Hold algorithm that applies a long or short position regarding the model price prediction. We also measure the risk of the strategy with the Sharp ratio.

For visualizing the results of the experiment, we use Tableau software to export graphs and tables based on CSV results from the pipeline Python implementation.

\subsection{Experiment 1 - processing model evaluation}
The first stage of experiment one is to evaluate which processing configuration will resolve with the best accuracy, F1 score, profit, and low risk. 

\subsubsection{Processing configuration evaluation\newline}
\begin{figure}[h]
\centering\includegraphics[width=0.7\linewidth]{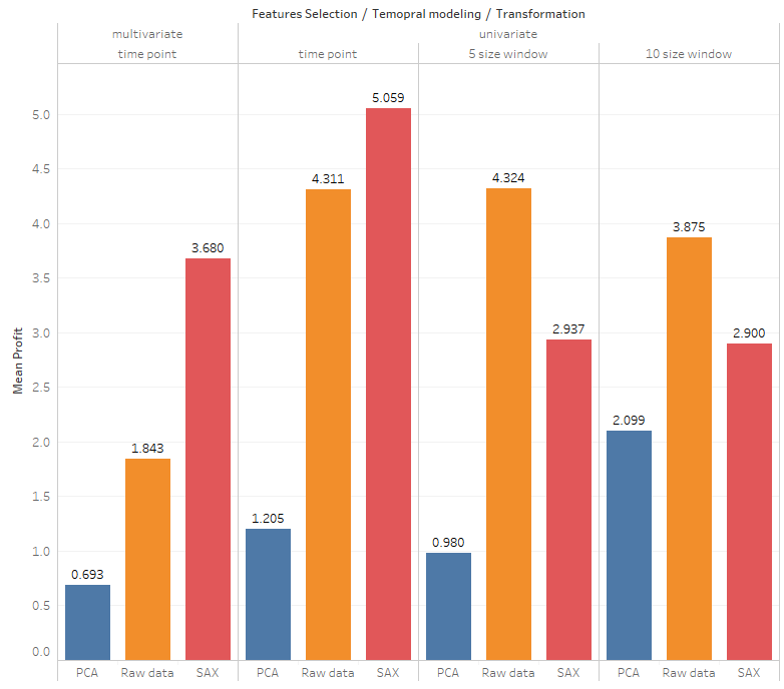}
\caption{Experiment 1 processing parameters results - mean profit values per transformation configuration
}
\label{fig:exp1_prep2}
\end{figure}

In figure ~\ref{fig:config}, we present the different configuration and metrics of the data processing configuration, and in figure ~\ref{fig:exp1_prep2}, we emphasize the profit difference per configuration. The best overall performance configurations are univariate modeling with SAX transformation. Furthermore, SAX transformation showed the best results with all other configurations. Each metric in the figure is the mean of 1680 examples (7 different stocks, five-folds, three different horizons, two predictions values, four types of models and two types of K top stocks)

\subsubsection{Prediction models evaluation\newline}
\begin{figure}[h]

\centering\includegraphics[width=1\linewidth]{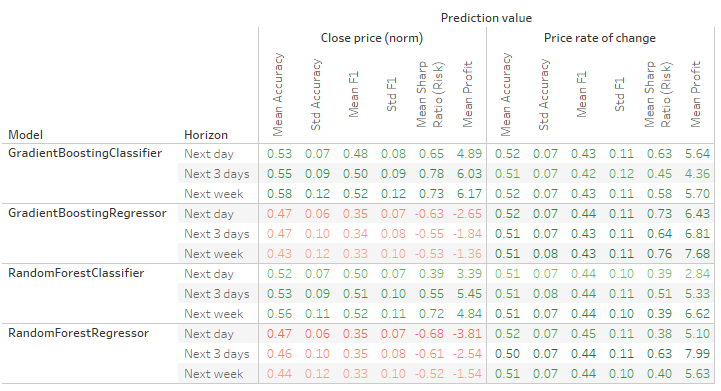}
\caption{Experiment 1 prediction parameters results - prediction model, Horizon and Value (rows - configuration, columns - prediction value with metrics and color - profit scale)
}
\label{fig:exp1_models}
\end{figure}

In the next step of the experiment, we evaluate the parameters of the prediction, from the results in figure ~\ref{fig:exp1_models} we can observe that the overall performance for predicting rate of change price (PROC) is higher than predicting the closing price (the columns of price rate of change is all green with only positive mean profits). 

The classification models have good accuracy results, but their standard deviation is high and results with negative profit along with high risk. The performance of the next day prediction horizon is higher than other horizons. For model performance, we witness an interesting behavior between the regressors and the classifiers. The classifiers had the best accuracy for predicting the closing price, whereas the regressors failed. However, the classifiers also meet negative mean profit, and the regressors did not, probably because of the prediction inconsistency (high standard deviation) and threshold calibration. From the results, we see a slight but not significant advantage of the Gradient boosting trees over the Random forest models.

\begin{figure}[h]
\centering\includegraphics[width=0.9\linewidth]{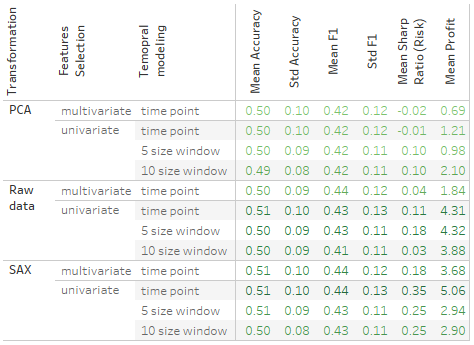}
\caption{Experiment 1 processing parameters results - transformation function, features and temporal modeling. (rows - configuration , columns - metrics and color - profit scale)
}
\label{fig:exp1_prep1}
\end{figure}

As a part of experiment one, we also evaluate the models that train on top 10 similar stocks calculated with simple Euclidean distance alongside models that trained only on the target stock. In figure ~\ref{fig:exp1_sim}, the results point out that the model trained only on the target stock has better results than the model trained on the ten similar stocks using Euclidean distance. 

From this experiment results, we conclude that the best processing and prediction parameters for the next experiment will be SAX transformation only on the price rate of change value (univariate). The price rate of change will also be the prediction value. 

We did not witness any significant results for the temporal modeling and prediction models and horizon prediction configurations; therefore, in the next experiment will apply all these configurations as well.  

\subsection{Experiment 2 - Similarity enhancement evaluation}
The second experiment train the models with the processing suggested from the first experiment and apply the similarity enhancement configurations as follows: similarity function, similarity value, top k stock to choose and the stock length fixing.

In this section, we present the evaluation figures only on profit. We found the accuracy and profit are correlated with each other. The full results are in appendix A.

\subsubsection{Similarity functions evaluation\newline}
\begin{figure}[h]
\centering\includegraphics[width=0.5\linewidth]{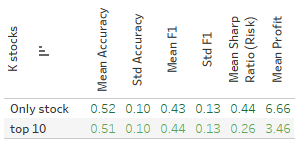}
\caption{Experiment 1 basic similarity results - a comparison between a model with top 10 similar stocks with Euclidean distance and a model without similarity enhancement. for predictions parameters: prediction model, Horizon and Value (rows - configuration, columns - metrics and color - profit scale)
}
\label{fig:exp1_sim}
\end{figure}

\begin{figure}[h]
\centering\includegraphics[width=0.8\linewidth]{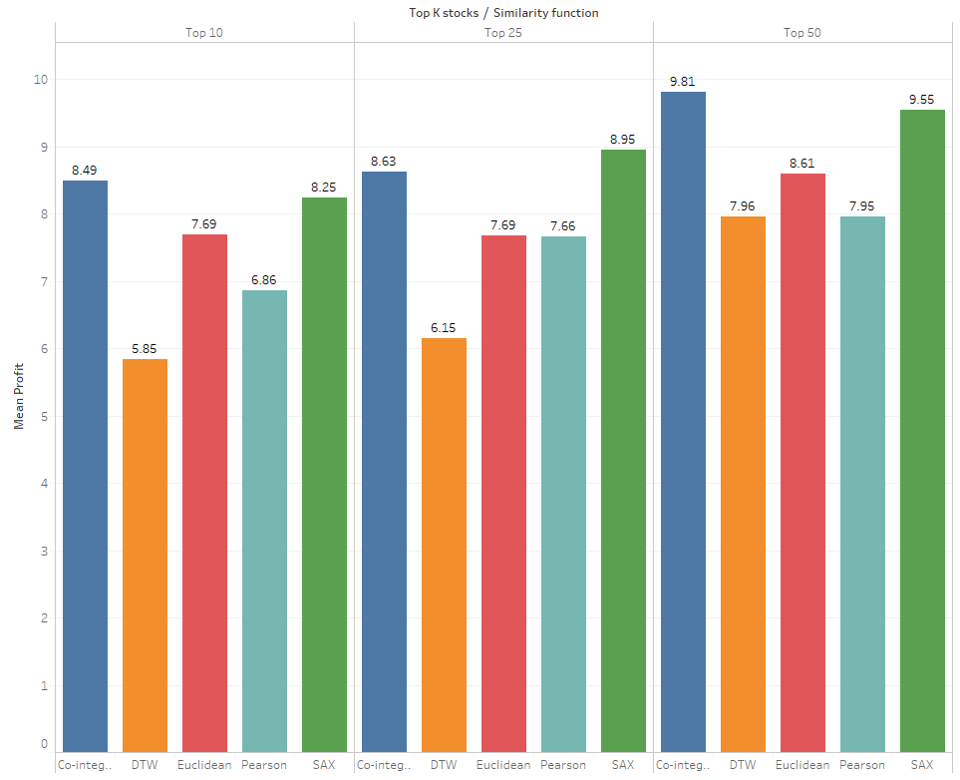}
\caption{Experiment 2 similarity configurations - a profit comparison between similarity configurations
}
\label{fig:exp2_sims2}
\end{figure}

The results from figure ~\ref{fig:exp2_sims2} map the metrics for top K stock and similarity stocks (rows) over the similarity value used for computing the similarity. Each cell in the table is a mean of 1680 instances (seven stocks, five-folds, three horizons, four models, 4 fixing length techniques). The results present a clear advantage of the co-integration and the SAX MINDIST similarities, price rate of change as similarity value, and selecting the top 50 stock. These combinations lead to a mean accuracy of 0.53 and a mean profit of more than 9.55 (SAX) and 9.81 (co-integration).  The results already show a significant improvement from the baseline model presented in figure ~\ref{fig:exp1_sim}.

\begin{figure}[H]
\begin{subfigure}[b]{0.8\textwidth}
\centering\includegraphics[width=0.8\linewidth]{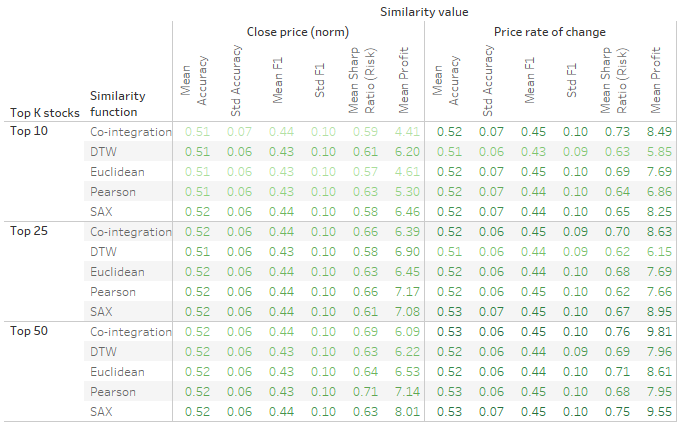}
\caption{Experiment 2 similarity configurations - a full metrics comparison between similarity configurations
}
\label{fig:exp2_sims}
\end{subfigure}

\begin{subfigure}[h]{0.4\textwidth}
\centering\includegraphics[width=0.8\linewidth]{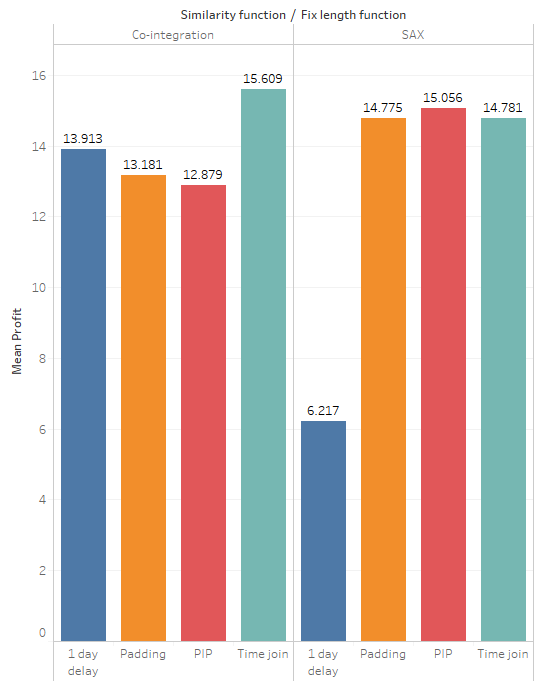}
\caption{Experiment 2 length fixing functions - a profit comparison between different fixing functions on top similarity configurations
}
\label{fig:exp2_fix}
\end{subfigure}

\end{figure}

In figure ~\ref{fig:exp2_fix}, we evaluate the length fixing functions of the best similarity configuration: the 50 top stocks with high SAX or co-integration similarity on price rate of change. Time join for fixing have the best results with a profit of 15.67 (co-integration) and 14.85 (SAX). 

In figure ~\ref{fig:exp2_rand_sim}, we present the final results of these similarities with the best performing processing and modeling configurations: a gradient boosting regressor trained on the price rate of change with SAX transformation and 50 top stocks with co-integration similarity. 

\subsubsection{Random stock enhancement comparison\newline}
\begin{figure}[h]
\centering\includegraphics[width=0.8\linewidth]{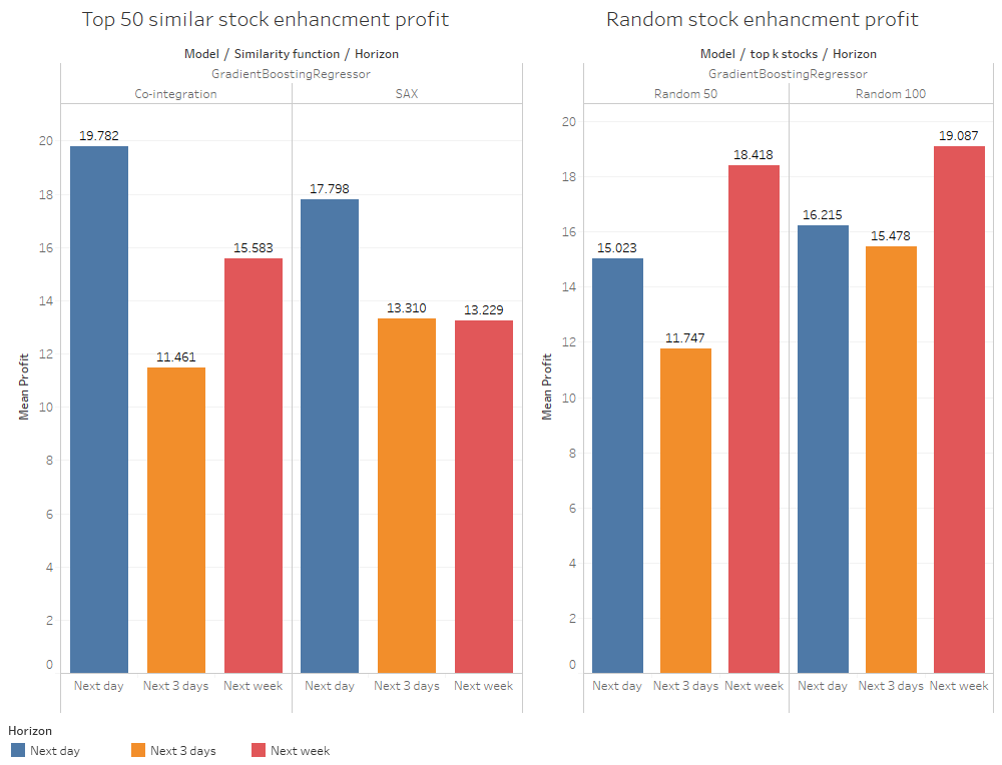}
\caption{Experiment 2 random selection compare - a profit comparison between SAX and co-integration similarities on top 50 stocks and random stock selection
}
\label{fig:exp2_rand_sim}
\end{figure}

Finally, we compare the results from the enhanced model with models that we enhanced with random 50 and 100 stocks. As described in ~\ref{fig:exp2_rand_sim}, the random enhanced model also had significantly better results than the baseline model presented in the first experiment. The random models improve as the horizon rise, and the number of random stocks is selected, this phenomena can be explained by the fact that the S\&P stocks are known to have similar behavior and can contribute the predictions because many investors and ETFs buy or sell the index stocks all together causing the prices to behave similarly. 

For the horizon of the next day, the co-integration stock similarity has significantly higher profit from the random 100 stocks with 19.78 and 16.21. On the other hand, the random model is significantly more profitable in the long horizons. The long horizon performance is a result of the same phenomena explained above regarding the S\&P stocks. Nevertheless, the co-integration based model is more accurate in terms of accuracy score and F1 score than the random 100 stocks model with accuracy between 0.542 - 0.55 and an F1 score between 0.448 - 0.459. The random 100 stocks model had less accurate results with an accuracy score between 0.526 - 0.535 and an F1 score between 0.437 - 0.443 (these detailed results are in appendix A). 

\begin{figure*}[h]
\centering\includegraphics[width=0.7\linewidth]{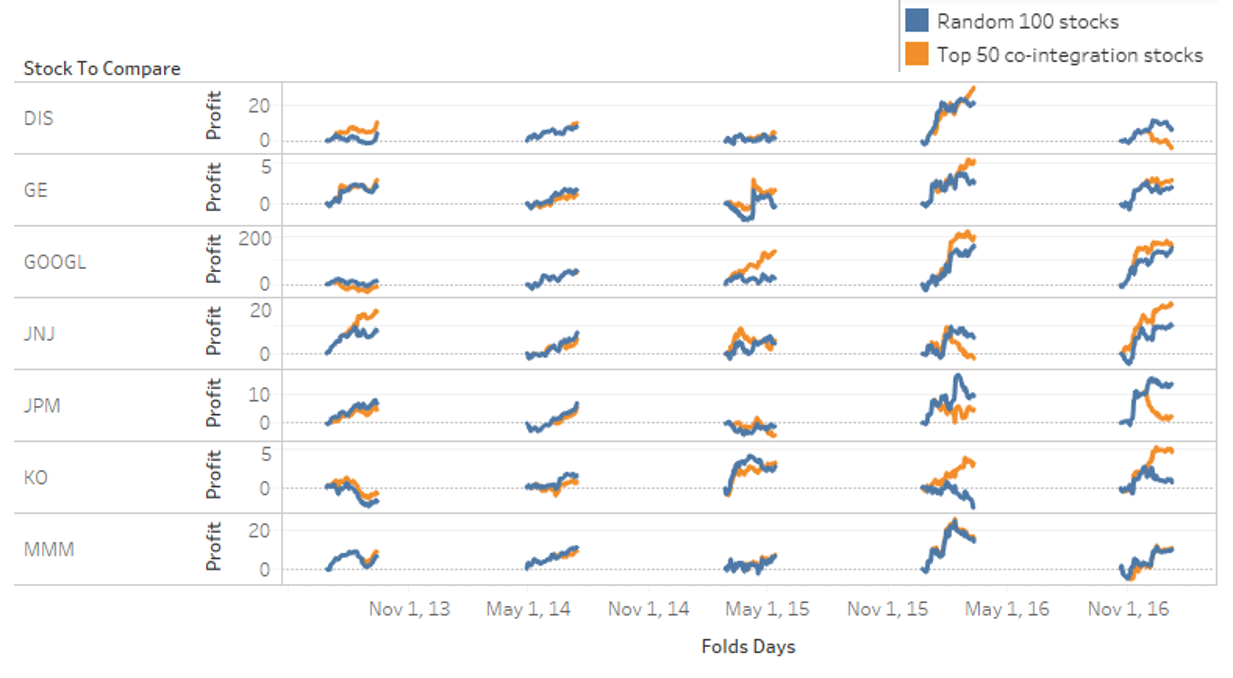}
\caption{Experiment 2 folds profit per stock - a profit comparison between top 50 stocks from co-integration similarity (orange) and 100 random stock selection enhancement (Blue) for each stock (x axis) in different folds (y axis)
}
\label{fig:exp2_profit_plot}
\end{figure*}

We further investigate the profit behavior of the co-integration model and the random 100 stocks model in order to understand the model's profitability behavior. In figure ~\ref{fig:exp2_profit_plot} we plot the profit value over time for each stock (x-axis) in each of the five folds evaluated (y-axis), the models predict the next day value and then the simple buy and hold strategy is applied, the color represents each of the two models. From the plots, we identify the co-integration model (orange color) to be more profitable in most stock's folds except JPM stock. 

\section{Conclusions}
In this paper, we focus on improving prediction models on stock data with similar stocks; the process of enhancement is not straightforward and requires several data processing phases. We design a pipeline for applying back-testing for all processing and prediction configurations. 

We came up with an optimized enhanced model with the following configurations: data processing of 10 size windows with the price rate of change SAX transformation. The predictor is a gradient boosting regressor with a 0.02 learning rate, and its training set is compound from the top 50 similar stocks found with co-integration similarity. We compared the enhanced model, the optimized baseline, and the random similarity model on seven stocks from different industries in five folds split over five years period. 

The enhanced model had better results than the other models in terms of accuracy and profit. The mean accuracy of the enhanced model is 0.55 compare the 0.52 and 0.546 of the non-enhanced and random enhanced models (respectively). In terms of profit, the enhanced model showed a high mean profit of 19.87 compared to 6.66 and 15.02 of the non-enhanced and random enhanced models.

During the research, we identify two limitations; the first regards the small data volume of the daily data set because the scope of daily prices is not enough data to train a well-fitted model. We believe that applying the pipeline on intra-day data might improve the models because of the data volume that may consist of much more useful similarities patterns.

The second limitation regards the S\&P stocks index in general. The index has two types of limitations; The first is that the index was mostly positive after the crisis of 2008; this behavior may affect the results in all the models evaluated. The other type is that the S\&P stocks correlate to each other because traders usually buy the entire S\&P index causing all the stocks to increase or decrease together. This kind of behavior eliminates some of the advantages a similarity measure might have because the stocks are already similar. In order to address these limitations, we aim to apply the pipeline on other investment instruments like crypt-currency and commodities (Gold, Oil, silver, etc.). We believe that the similarity pipeline on other datasets where most of the items do not correlate will have dramatically better results than in this research.  

For future work, we suggest some improvements. The most straightforward is to apply an ensemble similarity model on the different similarity measurements to combine the advantage of each method. Another improvement is to use deep learning models; we assume that the similarity enhancement will increase the training dataset and will improve deep learning models that require much more data for training.


\bibliographystyle{ACM-Reference-Format}
\bibliography{main} 

\pagebreak

\onecolumn
\section{Appendix A - similarity configuration Evaluations}

In this appendix, we collect the full evaluations per similarity configuration; these results address in the experiment results chapter.

\begin{figure}[H]
\begin{subfigure}[h]{0.6\textwidth}
\centering\includegraphics[width=1\linewidth]{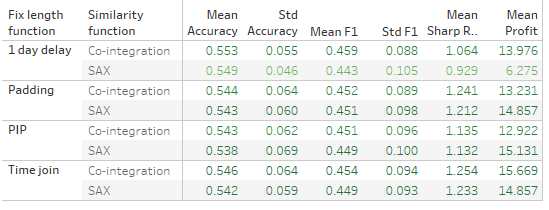}
\caption{Experiment 2 length fixing functions - a full metrics comparison between different fixing functions on top similarity configurations
}
\label{fig:exp2_t_fix}
\end{subfigure}

\begin{subfigure}[h]{0.6\textwidth}
\centering\includegraphics[width=1\linewidth]{figures/exp2_t_fix}
\caption{Experiment 2 length fixing functions - a full metrics comparison between different fixing functions on top similarity configurations
}
\label{fig:exp2_t_fix}
\end{subfigure}

\begin{subfigure}[h]{0.8\textwidth}
\centering\includegraphics[width=0.8\linewidth]{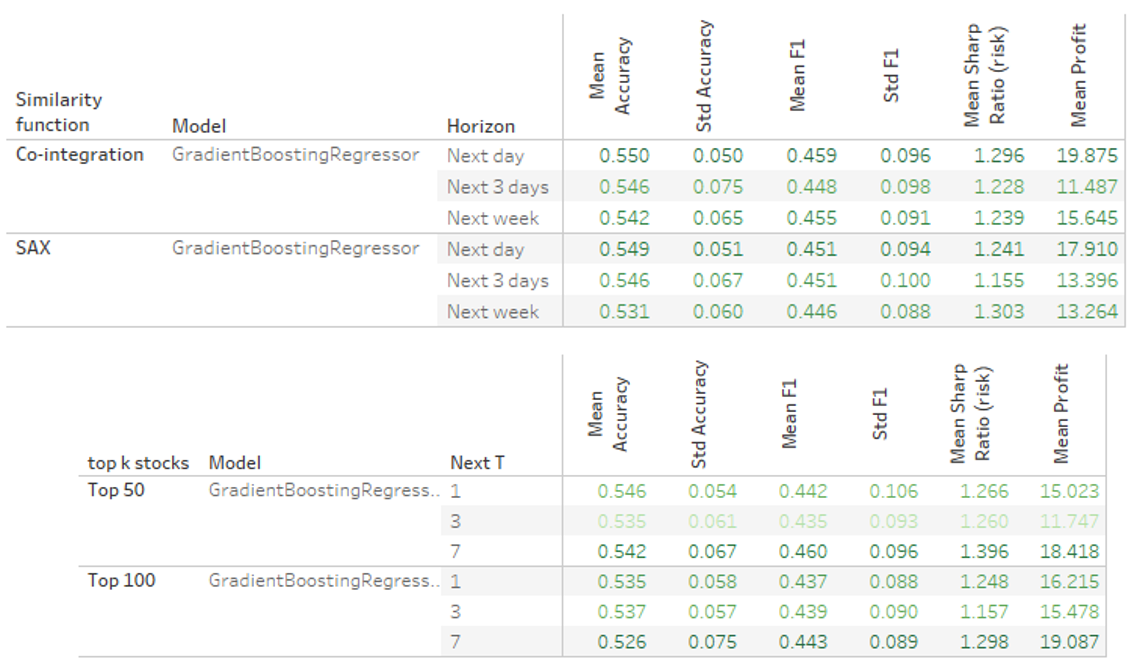}
\caption{Experiment 2 random selection compare - a full metrics comparison between SAX and co-integration similarities on top 50 stocks and random stock selection
}
\label{fig:exp2_t_randvs}
\end{subfigure}

\end{figure}

\pagebreak
\onecolumn
\section{Appendix B - stock similarity explorations}

In this appendix, we describe the manual exploration we conducted on the similarity results we conducted. In order to reason the similarity function, we tested if the similarity function can group stocks in the same industry.

\begin{figure}[h]
\centering\includegraphics[width=1\linewidth]{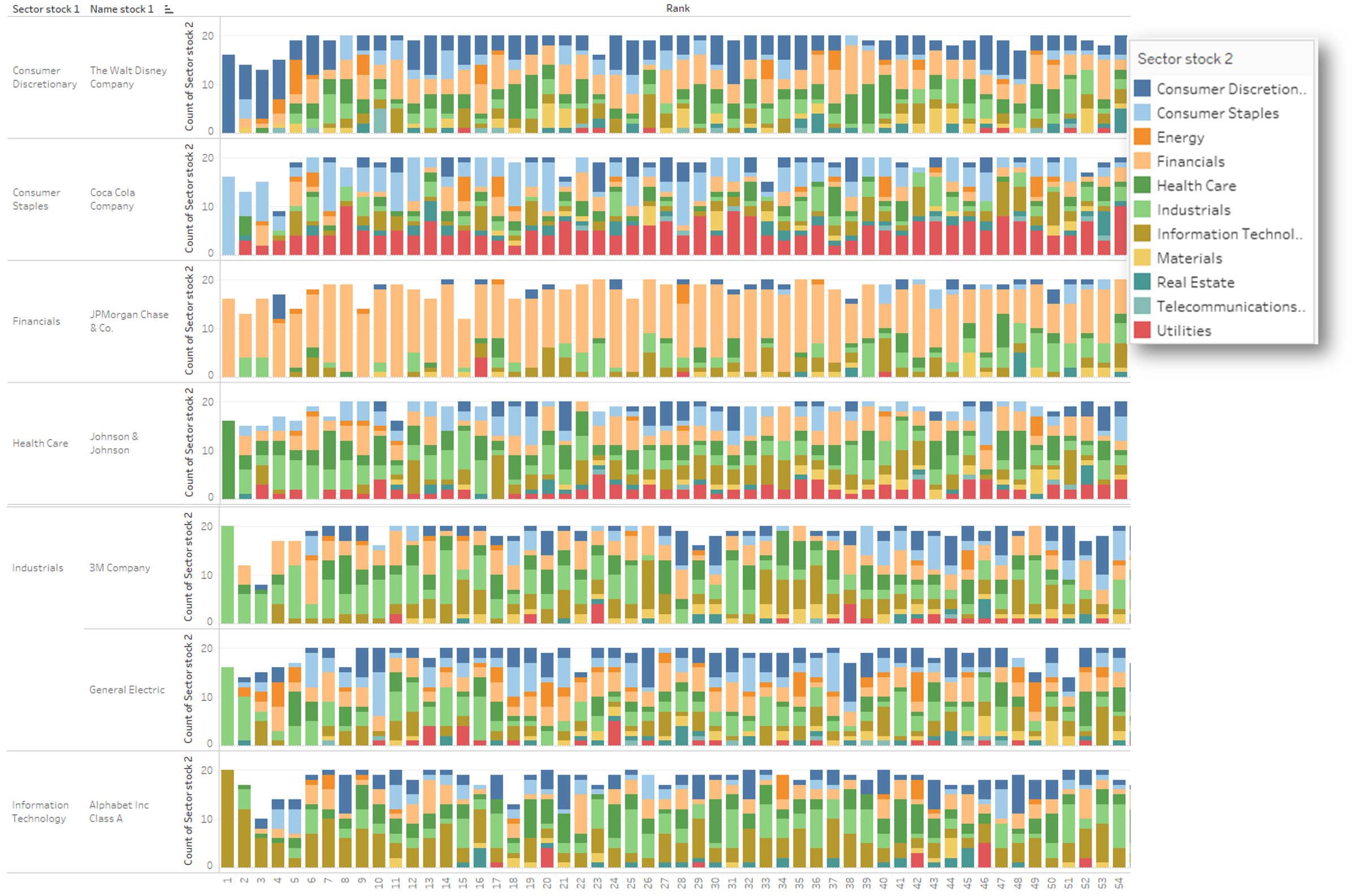}
\caption{Target stock sector spread per rank
}
\label{fig:sector1}
\end{figure}

In Figure ~\ref{fig:sector1} each row is a target stock, the bin graph contains the similar stocks found in a certain rank, and the color represents the sector count. For example, in Disney target function, the first bin is an all dark blue that represents the consumer sector, the same sector of the Disney stock. That means all the similarity functions choose the most similar stock from the same sector as Disney; If we continue with the y-axis, we can see how the target stock sector is dominant in the beginning and then spread, this means that the similarities find a relationship between the stocks' behavior and their sector.

\begin{figure}[h]
\centering\includegraphics[width=0.7\linewidth]{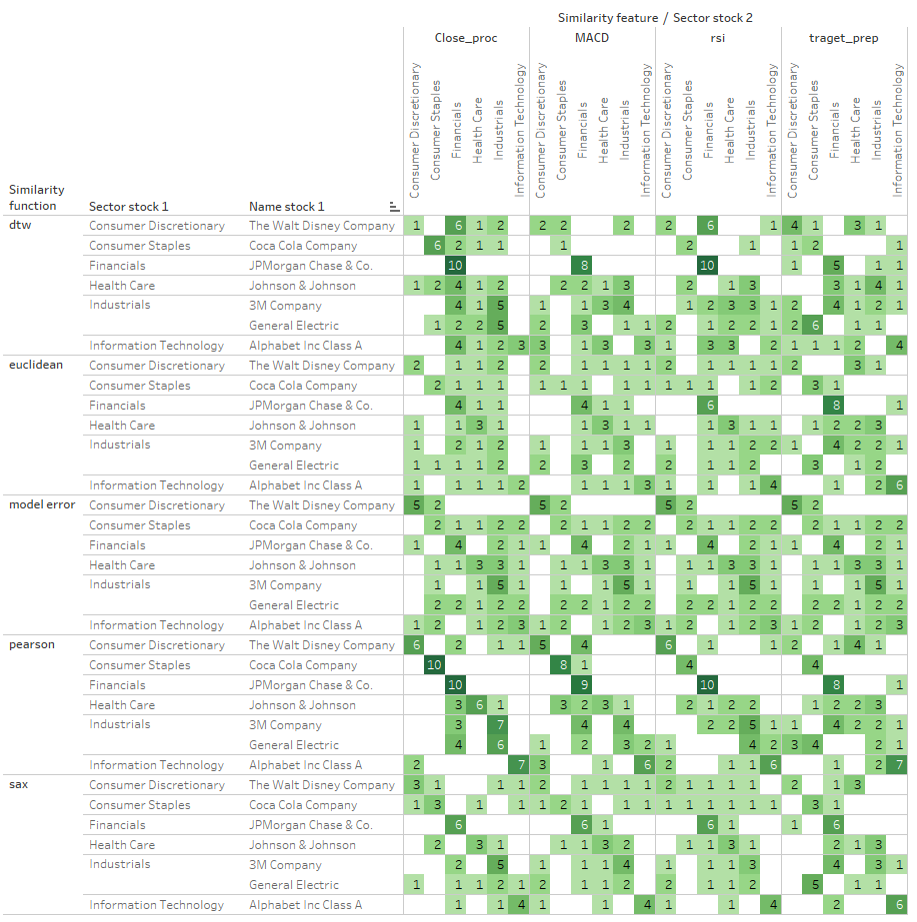}
\caption{Target stock and sector confusion matrix for top 10 stocks per similarity function (vertical) and similarity value (horizontal)
}
\label{fig:sector2}
\end{figure}

In Figure ~\ref{fig:sector2}, we demonstrate a confusion matrix between the target stock and the sector for the top 10 similar stocks found per similarity values (horizon) and similarity function (vertical). For example, the Pearson and close proc (Price rate of change) on the coca-cola stock found that all top 10 similar stocks belong to the same sector as coca-cola, consumer staples. The stocks and sectors arranged in a way that each stock corresponds to its relevant sector, meaning that if the diagonal is all dark green with ten value, the similarity function found the stocks similar to its sector's stocks. 

In Figure ~\ref{fig:sector2}, we describe which stocks found most similar to the target stocks, for example, in the coca-cola stocks, the similarities functions found Pepsi to be the most similar.

\begin{figure}[h]
\centering\includegraphics[width=1\linewidth]{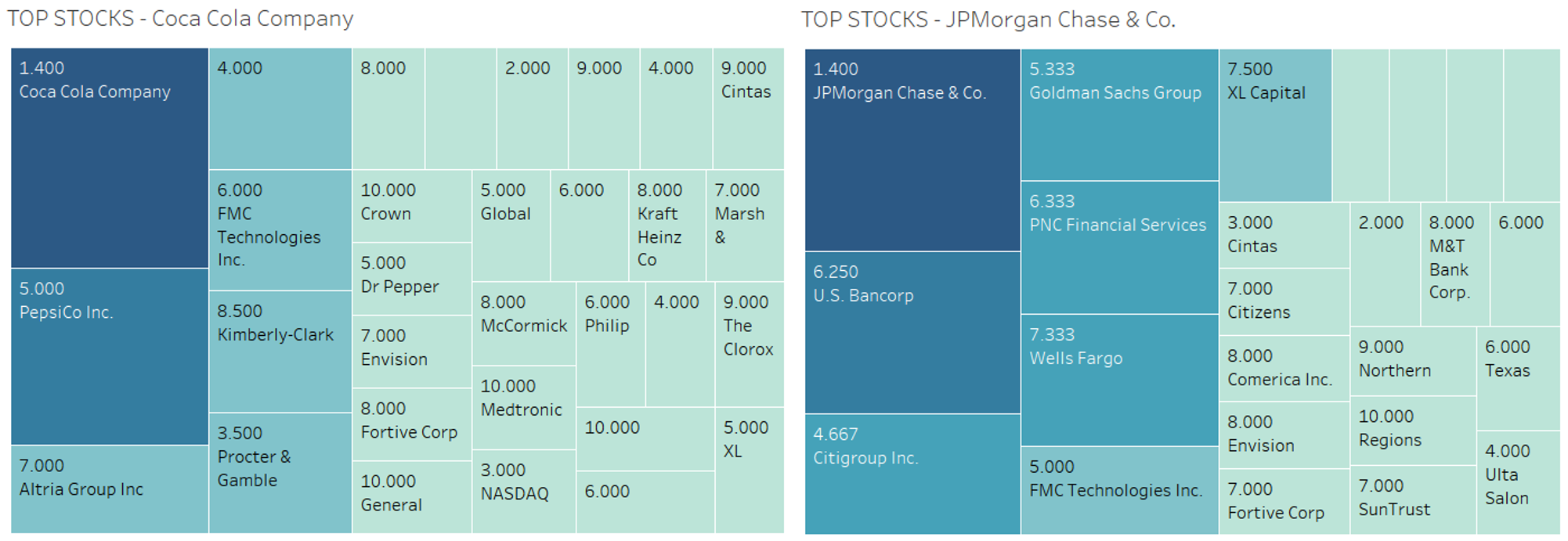}
\caption{Coca cola and JP Mrgan Top stocks chosen
}
\label{fig:sector3}
\end{figure}

\end{document}